\documentclass[a4,12pt]{article}
\usepackage{amsmath,amssymb}
\newcommand{\nc}{\newcommand}
\nc{\bg}{B. Grzadkowski}
\nc{\non}{\nonumber}
\def\dps{\displaystyle}
\def\mib#1{\mbox{\boldmath $#1$}}

\def\bra#1{\langle #1 |} \def\ket#1{|#1\rangle}
\def\vev#1{\langle #1\rangle}

\nc{\barx}{\bar{x}}\nc{\pbarn}{\;\hbox {pb}}\nc{\fbarn}{\;\hbox {fb}}
\nc{\hc}{\hbox {h.c.}} \nc{\re}{\hbox {Re}} 
\nc{\mev}{\hbox {MeV}} \nc{\gev}{\;\hbox {GeV}}
\def\gesim{\lower0.5ex\hbox{$\:\buildrel >\over\sim\:$}}
\def\lesim{\lower0.5ex\hbox{$\:\buildrel <\over\sim\:$}}
\nc{\prd}[3]{{\it Phys.\ Rev.}\ {{\bf D{#1}} (#2) #3}}
\nc{\prl}[3]{{\it Phys.\ Rev.\ Lett.}\ {{\bf {#1}} (#2) #3}}
\nc{\plb}[3]{{\it Phys.\ Lett.}\ {{\bf B{#1}} (#2) #3}}
\nc{\npb}[3]{{\it Nucl.\ Phys.}\ {{\bf B{#1}} (#2) #3}}
\nc{\ptp}[3]{{\it Prog.\ Theor.\ Phys.}\ {{\bf {#1}} (#2) #3}}
\nc{\zfp}[3]{{\it Z.\ Phys.}\ {{\bf C{#1}} (#2) #3}}
\nc{\epj}[3]{{\it Eur.\ Phys.\ J.}\ {{\bf C{#1}} (#2) #3}}
\nc{\mpla}[3]{{\it Mod.\ Phys.\ Lett.}\ {{\bf A{#1}} (#2) #3}}
\nc{\rmp}[3]{{\it Rev.\ Mod.\ Phys.}\ {{\bf {#1}} (#2) #3}}
\nc{\ijmpa}[3]{{\it Int.\ J.\ Mod.\ Phys.}\
               {{\bf A{#1}} (#2) #3}}
\nc{\ttbar}{t\bar{t}}         \nc{\bbbar}{b\bar{b}}
\nc{\tanb}{\tan \beta}        \nc{\twbdec}{t\to W^+ b}
\nc{\tbwbdec}{\bar{t}\to W^- \bar{b}}
\nc{\epem}{e^+e^-}            \nc{\eett}{\epem \to \ttbar}
\nc{\sigeett}{\sigma_{e\bar{e}\to\ttbar}}
\nc{\wpwm}{W^+W^-}            \nc{\tbar}{\bar{t}}
\nc{\bbar}{\bar{b}}           \nc{\wpp}{W^+}
\nc{\mt}{m_t}    \nc{\mts}{m_t^2}   \nc{\mw}{m_W}    \nc{\mws}{m_W^2}
\nc{\mz}{m_Z}    \nc{\mzs}{m_Z^2}
\nc{\ttbardec}{\ttbar \to W^+W^-\bbbar}
\nc{\wwbb}{W^+W^-\bbbar}      \nc{\sm}{SM}
\nc{\cw}{\cos\theta_W}        \nc{\sw}{\sin\theta_W}
\nc{\sws}{\sin^2\theta_W}     \nc{\sig}{\sigma_{tot}}
\nc{\lp}{{\ell}^+}              \nc{\lm}{{\ell}^-}
\nc{\epsl}{\epsilon_L}        \nc{\cp}{C\!P}
\nc{\gaga}{\gamma\gamma}
\nc{\splus}{s_+}       \nc{\smin}{s_-}        \nc{\eps}{\epsilon}
\nc{\psp}{Ps_+}        \nc{\psm}{Ps_-}        \nc{\lsp}{ls_+}
\nc{\lsm}{ls_-}        \nc{\sss}{s_+s_-}      \nc{\m}{m_t}
\nc{\mq}{m_t^2}        \nc{\mr}{\frac{1}{\m}} \nc{\av}{A_{\gamma}}
\nc{\bv}{B_{\gamma}}   \nc{\az}{A_Z}          \nc{\bz}{B_Z}
\nc{\avs}{A_{\gamma}^2}\nc{\azs}{A_Z^2}       \nc{\bzs}{B_Z^2}
\nc{\dav}{\delta \! A_{\gamma}}   \nc{\dbv}{\delta \! B_{\gamma}}
\nc{\dcv}{\delta C_{\gamma}}      \nc{\ddv}{\delta \! D_{\gamma}}
\nc{\daz}{\delta \! A_Z}          \nc{\dbz}{\delta \! B_Z}
\nc{\dcz}{\delta C_Z}             \nc{\ddz}{\delta \! D_Z}
\nc{\dev}{\delta \! E_{\gamma}}   \nc{\dez}{\delta \! E_Z}
\nc{\dfv}{\delta \! F_{\gamma}}   \nc{\dfz}{\delta \! F_Z}
\nc{\rdav}{{\rm Re}(\delta \! A_{\gamma}) \:}
\nc{\rdbv}{{\rm Re}(\delta \! B_{\gamma}) \:}
\nc{\rdcv}{{\rm Re}(\delta C_{\gamma}) \:}
\nc{\rddv}{{\rm Re}(\delta \! D_{\gamma}) \:}
\nc{\rdaz}{{\rm Re}(\delta \! A_Z) \:}
\nc{\rdbz}{{\rm Re}(\delta \! B_Z) \:}
\nc{\rdcz}{{\rm Re}(\delta C_Z) \:}
\nc{\rddz}{{\rm Re}(\delta \! D_Z) \:}
\nc{\idav}{{\rm Im}(\delta \! A_{\gamma}) \:}
\nc{\idbv}{{\rm Im}(\delta \! B_{\gamma}) \:}
\nc{\idcv}{{\rm Im}(\delta C_{\gamma}) \:}
\nc{\iddv}{{\rm Im}(\delta \! D_{\gamma}) \:}
\nc{\idaz}{{\rm Im}(\delta \! A_Z) \:}
\nc{\idbz}{{\rm Im}(\delta \! B_Z) \:}
\nc{\idcz}{{\rm Im}(\delta C_Z) \:}
\nc{\iddz}{{\rm Im}(\delta \! D_Z) \:}
\nc{\cz}{(1+v_e^2)d\:\!'^2}         \nc{\ci}{v_ed\:\!'}
\nc{\ccz}{v_ed\:\!'^2}              \nc{\cci}{d\:\!'}
\nc{\lspace}{\;\;\;\;\;\;\;\;\;\;}  \nc{\llspace}{\lspace \lspace}
\nc{\beq}{\begin{equation}}   \nc{\eeq}{\end{equation}}
\nc{\bea}{\begin{eqnarray}}   \nc{\eea}{\end{eqnarray}}
\nc{\baa}{\begin{array}}      \nc{\eaa}{\end{array}}
\nc{\bit}{\begin{itemize}}    \nc{\eit}{\end{itemize}}
\nc{\ben}{\begin{enumerate}}  \nc{\een}{\end{enumerate}}
\nc{\bce}{\begin{center}}     \nc{\ece}{\end{center}}
\nc{\ocal}{{\cal O}}
\newcounter{QQ}
\newcounter{QQF}
\newcounter{QQE}
\begin{document}
\pagestyle{empty} \setlength{\footskip}{2.0cm}
\setlength{\oddsidemargin}{0.5cm} \setlength{\evensidemargin}{0.5cm}
\renewcommand{\thepage}{-- \arabic{page} --}
\def\mib#1{\mbox{\boldmath $#1$}}
\def\bra#1{\langle #1 |}      \def\ket#1{|#1\rangle}
\def\vev#1{\langle #1\rangle} \def\dps{\displaystyle}
\nc{\tb}{\stackrel{{\scriptscriptstyle (-)}}{t}}
\nc{\bb}{\stackrel{{\scriptscriptstyle (-)}}{b}}
\nc{\fb}{\stackrel{{\scriptscriptstyle (-)}}{f}}
\nc{\pp}{\gamma \gamma}
\nc{\pptt}{\pp \to \ttbar}
   \def\thebibliography#1{\centerline{REFERENCES}
     \list{[\arabic{enumi}]}{\settowidth\labelwidth{[#1]}\leftmargin
     \labelwidth\advance\leftmargin\labelsep\usecounter{enumi}}
     \def\newblock{\hskip .11em plus .33em minus -.07em}\sloppy
     \clubpenalty4000\widowpenalty4000\sfcode`\.=1000\relax}\let
     \endthebibliography=\endlist
   \def\sec#1{\addtocounter{section}{1}\section*{\hspace*{-0.72cm}
     \normalsize\bf\arabic{section}.$\;\;$#1}\vspace*{-0.3cm}}

   \def\subsec#1{\addtocounter{subsection}{1}\subsection*{\hspace*{-0.72cm}
    \normalsize\bf\arabic{section}.\arabic{subsection}.$\;\;$#1}
    \vspace*{-0.3cm}}
\vspace*{-1.0cm}
\noindent
\pagestyle{empty} \setcounter{page}{0}
 \vspace{-0.7cm}
\begin{flushright}
$\vcenter{
\hbox{{\footnotesize IFT-03-06~~~~FUT-06-01}}
\hbox{{\footnotesize UCRHEP-T404}}
\hbox{{\footnotesize TOKUSHIMA Report}}
\hbox{(hep-ph/0602130)}
}$
\end{flushright}
\renewcommand{\thefootnote}{$\dag$}
\vskip 0.8cm
\begin{center}
{\large\bf New-Physics Search through 
$\boldsymbol{\gamma\gamma\to t\bar{t}\to\ell X /b X}$}~\footnote{Presented by
 K. Ohkuma 
at the 7th International Symposium 
on Radiative Corrections (RADCOR2005), Shonan Village, 
Japan, October 2-7, 2005.\\
E-mail address: \tt ohkuma@fukui-ut.ac.jp}
\end{center}

\vspace{0.4cm}
\begin{center}
\renewcommand{\thefootnote}{\alph{footnote})}
{\sc Bohdan GRZADKOWSKI$^{\:1) \:}$}, \  
{\sc Zenr\=o HIOKI$^{\:2) \:}$}

\vskip 0.15cm
{\sc Kazumasa OHKUMA$^{\:3) \:}$}\ 
\ and \ \ 
{\sc Jos\'e WUDKA$^{\:4) \:}$}
\end{center}

\vspace*{0.2cm}
\centerline{\sl $1)$ Institute of Theoretical Physics,\ Warsaw
University}
\centerline{\sl Ho\.za 69, PL-00-681 Warsaw, Poland}

\vskip 0.2cm
\centerline{\sl $2)$ Institute of Theoretical Physics,\
University of Tokushima}
\centerline{\sl Tokushima 770-8502, Japan}

\vskip 0.2cm
\centerline{\sl $3)$ Department of Information Science,\
Fukui University of Technology}
\centerline{\sl Fukui 910-8505, Japan}

\vskip 0.2cm
\centerline{\sl $4)$ Department of Physics,\
University of California Riverside}
\centerline{\sl Riverside, CA 92521-0413, USA}

\vspace*{1.5cm}
\centerline{ABSTRACT}

\vspace*{0.3cm}
\baselineskip=20pt plus 0.1pt minus 0.1pt
We probe optimal beam polarizations for new-physics search
in top-quark and Higgs-boson sectors at Photon Linear Colliders
(PLC). Expressing possible non-standard effects generated by
$SU(2) \times U(1)$ gauge-invariant dimension-6 effective
operators as anomalous top and Higgs couplings, we estimate
expected statistical sensitivities of these couplings in
$\gamma \gamma \to t\bar{t} \to \ell X / b X$, using the
optimal-observable method.
\vspace*{0.4cm} \vfill



\newpage
\renewcommand{\thefootnote}{$\sharp$\arabic{footnote}}
\pagestyle{plain} \setcounter{footnote}{0}
\pagestyle{plain} \setcounter{page}{1}
\baselineskip=21.0pt plus 0.2pt minus 0.1pt
\sec{INTRODUCTION}
Recent neutrino experiments strongly indicate that the Standard Model
(SM) has to be modified to explain $\nu$-oscillation phenomena
\cite{Langacker:2005pf}. However this does not necessarily require
any new particles and we have no information yet on SM behavior in
much higher-energy region where even new heavy particles could produce
some effects directly or indirectly. Since top-quark/Higgs-boson
interactions might receive corrections via such new-physics effects,
precise studies of their reactions are expected to be a window on
possible beyond-the-SM physics.

Current measurements of top quark at Fermilab Tevatron, which is the
unique facility for top-quark production at present, are not sufficient
yet to study if top-quark can be fully described in the framework of
SM \cite{TopTev}, but in the near future Large Hadron Collider
(LHC) \cite{LHC} and International Linear Collider (ILC) \cite{ILC}
will work as top-quark factories and realize its precise studies.
Considering that the top-Higgs coupling is proportional to $m_t$,
these colliders will also give us an opportunity to study the
Higgs sector.

Focusing on ILC, we have been carrying out those studies intensively.
Here we would like to show some of our results about drawing possible
anomalous top and Higgs couplings from the processes $\gamma \gamma
\to t\bar{t}\to\ell/b X$ at Photon Linear Collider (PLC), an interesting
option of ILC, and discuss optimal beam polarizations that could lead
us to the smallest statistical uncertainties. This report is organized
as follows; We briefly review the framework of the analysis in
section 2. Our numerical results are presented in section 3, and then
a summary and some discussions are contained in section.4.

\sec{FRAMEWORK}

In this section, we briefly explain the effective-Lagrangian
approach \cite{Buchmuller:1986jz} and optimal-observable
method \cite{optimal} used in our analysis.
In addition, polarization parameters of PLC are reviewed.

\subsec{Effective Lagrangian}

We assumed that all non-standard particles are heavier than
the lowest new-physics scale ${\mit \Lambda}$ and decouple
in lower-energy processes. In the effective-Lagrangian
approach with this assumption, effects from such heavy
particles can be parameterized as coupling constants of
effective operators which are composed of standard fields
alone.

Let us describe this scenario more concretely. The SM
Lagrangian ${\cal L}_{\rm SM}$ is modified by the
addition of a series of $SU(3)\times SU(2)\times U(1)$
gauge-invariant operators with coefficients suppressed by
inverse powers of ${\mit \Lambda}$. Among those
operators the largest contribution comes from dimension-6
operators, denoted as ${\cal O}_i$, since dimension-5
operators violate lepton number \cite{Buchmuller:1986jz}
and are irrelevant for the processes considered here.
Consequently we have
\begin{equation}
{\cal L}_{\rm eff}={\cal L}_{\rm SM}
+\frac1{{\mit\Lambda}^2}\sum_i (\alpha_i {\cal O}_i
+ {\rm h.c.}) + O( {\mit\Lambda}^{-3} )
\end{equation}
as the basis of our analysis.

The operators relevant here lead to the
following non-standard top-quark- and 
Higgs-boson-couplings:
(i) $C\!P$-conserving and $C\!P$-violating $t\bar{t}\gamma$ vertices,
(ii) $C\!P$-conserving and $C\!P$-violating $\gamma\gamma H$ vertices, and
(iii) the anomalous $tbW$ vertex. The corresponding coupling
constants are denoted respectively as
$\alpha_{\gamma 1}$, $\alpha_{\gamma 2}$, $\alpha_{h1}$,
$\alpha_{h2}$ and $\alpha_{d}$.
The explicit expressions for
these couplings in terms of the coefficients
of dimension-6 operators are to be found in 
Ref.~\cite{Grzadkowski:2003tf,Grzadkowski:2005ye}.

\subsec{Optimal-observable method}

The optimal-observable method \cite{optimal} is a useful tool
for estimating expected statistical uncertainties in various
coupling measurements, which we summarize below.

Suppose we have a cross section
\begin{equation}
\frac{d\sigma}{d\phi}(\equiv{\mit\Sigma}(\phi))=\sum_i c_i f_i(\phi),
\end{equation}
where $f_i(\phi)$ are known functions of the final-state
phase-space variables $\phi$ and $c_i$'s are model-dependent coefficients. 
The goal is to determine the $c_i$'s. This can be done by using
appropriate weighting functions $w_i(\phi)$ such that $\int w_i(\phi)
{\mit\Sigma}(\phi)d\phi=c_i$. In general, different choices for
$w_i(\phi)$ are possible, but there is a unique choice for which the
resultant statistical error is minimized. Such functions are given by
\begin{equation}
w_i(\phi)=\sum_j X_{ij}f_j(\phi)/{\mit\Sigma}(\phi)\,, \label{X_def}
\end{equation}
where $X_{ij}$ is the inverse matrix of ${\cal M}_{ij}$ which
is defined as
\begin{equation}
{\cal M}_{ij}
\equiv \int {f_i(\phi)f_j(\phi)\over{\mit\Sigma}(\phi)}d\phi\,.
\label{M_def}
\end{equation}
When we use these weighting functions, the statistical uncertainty
of $c_i$ becomes
\begin{equation}
{\mit\Delta}c_i=\sqrt{X_{ii}\,\sigma_T/N}\,, \label{delc_i}
\end{equation}
where $\sigma_T\equiv\int (d\sigma/d\phi) d\phi$ and $N$ is the total
number of events.

In order to adapt this method to our processes, we decompose
cross sections into a sum of several functions which are proportional
to each non-standard couplings\footnote{
    We kept only linear terms in non-standard couplings. 
}.
In this way, the angular and energy distributions of the secondary
fermions $\ell/b$ in the $e\bar{e}$ CM frame are expressed as
\begin{equation}
\frac{d\sigma}{dE_{\ell/b} d\cos\theta_{\ell/b}}
=f_{\rm SM}(E_{\ell/b}, \cos\theta_{\ell/b}) 
+ \sum_i \alpha_i f_i (E_{\ell/b}, \cos\theta_{\ell/b}),
\label{distribution}
\end{equation}
where $f_{\rm SM}$ and $f_i$ are calculable functions:
$f_{\rm SM}$ denotes the SM contribution,
$f_{\gamma 1,\gamma 2}$ describe the anomalous
$C\!P$-conserving and $C\!P$-violating
$t\bar{t}\gamma$-ver\-ti\-ces contributions respectively,
$f_{h1,h2}$ those generated by the anomalous $C\!P$-conserving
and $C\!P$-violating $\gamma\gamma H$-ver\-ti\-ces,
and $f_d$ that by the anomalous $tbW$-vertex (see, e.g.,
Ref.\cite{Grzadkowski:2003tf} for detailed calculations).

\subsec{Polarization parameters of PLC}

At PLC, a colliding photon beam originates 
as a laser beam back-scattered off an electron ($e$) or 
positron ($\bar{e}$) beam. 
The polarizations of the initial state are 
characterized by the electron and positron longitudinal 
polarizations $P_e$
and $P_{\bar{e}}$, the maximum average linear polarizations
$P_t$ and $P_{\tilde{t}}$ of the laser photons with
the azimuthal angles $\varphi$ and $\tilde{\varphi}$
(defined in the same way as in \cite{Ginzburg:1981vm,Borden:1992qd}),
and their average helicities $P_{\gamma}$ and $P_{\tilde{\gamma}}$.

The photon polarizations $P_{t,\gamma}$ and
$P_{\tilde{t},\tilde{\gamma}}$ satisfy
\begin{equation}
0 \leq P_t^2 + P_{\gamma}^2 \leq 1,
\ \ \ \ \ \ \ \ \
0 \leq P_{\tilde{t}}^2 + P_{\tilde{\gamma}}^2 \leq 1.
\end{equation}
For the linear polarization, we fixed the relative azimuthal
angle $\chi \equiv \varphi-\tilde{\varphi}$ to be $\pi/4$, because
we found it the optimal value to study $\alpha_{\gamma2}$ and
$\alpha_{h2}$ through checking $\chi$ dependence of
$\sigma(\gamma\gamma\to t\bar{t})$.

\sec{NUMERICAL RESULTS}
We searched for the polarizations that
could make the statistical uncertainties ${\mit\Delta} \alpha_i$
minimum for $\sqrt{s_{e\bar{e}}}=500$ GeV and ${\mit\Lambda}=1$
TeV, varying their parameters as
$P_{e,\bar{e}}=0,\,\pm1$, $P_{t,\tilde{t}}=0,
\,1/\sqrt{2},\,$ 1, and $P_{\gamma,\tilde{\gamma}}=0,
\,\pm 1/\sqrt{2},\,\pm1$.
 We also changed the Higgs mass as $m_H=$100, 300 and 500 GeV,
which lead to the width ${\mit\Gamma}_H=1.08\times 10^{-2}$, 8.38
and 73.4 GeV respectively according to the standard-model formula.

We did not probe the Higgs-resonance region, which had been
extensively studied previously (see, for example,
\cite{Gounaris:1997ef}). This means that the non-standard
corrections are expected to be moderate and we may compute
the number of secondary fermions, $N_{\ell/b}$, from the SM
total cross section
multiplied by the lepton/$b$-quark detection efficiency
$\epsilon_{\ell/b}$ and the integrated $e\bar{e}$ luminosity
$L_{e\bar{e}}\,$; this leads to $N_{\ell/b}$ independent of
$m_H$.

Before showing the corresponding results, we have to explain
one serious problem we encountered in Ref.\cite{Grzadkowski:2003tf}.
We have noticed that the numerical results for $X_{ij}$ are often
unstable when inverting the matrix ${\cal M}_{ij}$: even a tiny
variation of ${\cal M}_{ij}$
changes $X_{ij}$ significantly. This indicates that some of
$f_i$ have similar shapes and therefore their coefficients
cannot be disentangled easily.
The presence of such instability has forced us to refrain from
determining all the couplings at once through this process alone.
That is, we have assumed that some of $\alpha_i$'s had been
measured in other processes (e.g., in
$e\bar{e}\to t\bar{t}\to{\ell}^{\pm}X$), and we performed an
analysis with smaller number of independent parameters.

When estimating the statistical uncertainty in simultaneous
measurements, e.g., of $\alpha_{\gamma 1}$ and $\alpha_{h 1}$
(assuming all other coefficients are known), we need only
the components with indices 1, 2 and 4. In such a ``reduced
analysis'', we still encountered the instability problem, and
we selected ``stable solutions'' according to the following
criterion: Let us express the resultant uncertainties as
${\mit\Delta}\alpha_{\gamma 1}^{[3]}$ and
${\mit\Delta}\alpha_{h 1}^{[3]}$, where ``3'' shows that we
use the input ${\cal M}_{ij}$, keeping three decimal places.
In addition, we also compute
${\mit\Delta}\alpha_{\gamma 1}^{[2]}$
and ${\mit\Delta}\alpha_{h 1}^{[2]}$
by rounding ${\cal M}_{ij}$
off to two decimal places. Then, we accept the result as a
stable solution if both of the deviations
$|{\mit\Delta}\alpha_{\gamma 1,h 1}^{[3]}
-{\mit\Delta}\alpha_{\gamma 1,h 1}^{[2]}|/
{\mit\Delta}\alpha_{\gamma 1,h 1}^{[3]}$ are less than 10 \%.

This is what we faced in \cite{Grzadkowski:2003tf}, and we
found that we are not free from this problem even for the
wider range of polarization parameters given at the beginning
of this section. That is, we did not find again any stable
solution in the four- and five-parameter analysis. Fortunately,
however, we did find some solutions not only
in the two- but also in the three-parameter analysis
\cite{Grzadkowski:2005ye}. 
In this report, we show the results of two parameter analysis
as examples below. 
We did not fix the detection
efficiencies $\epsilon_{\ell/b}$ since they depend
on detector parameters and will get better with development of
detection technology.
On the other hand, the $N_{\ell/b}$ were estimated 
for $L_{e\bar{e}}=500$ fb${}^{-1}$.\vspace*{0.5cm}

\setcounter{QQ}{\arabic{equation}} \addtocounter{QQ}{1}
\setcounter{QQF}{\arabic{QQ}} 
\noindent
\underline{\underline{Final charged-lepton detection}}\vspace*{-0.35cm}\\
\begin{description}
\item[] \underline{Independent of $m_H$}
%
  \item[$\bullet$]
    $P_e=P_{\bar{e}}=-1, ~P_t = P_{\tilde{t}}=1,~
     P_\gamma = P_{\tilde{\gamma}}=0$,~
     $N_{\ell}\simeq 1.0\times 10^4 \epsilon_{\ell}$\\
     ${\mit\Delta} \alpha_{\gamma1}=0.051/\sqrt{\epsilon_{\ell}}$,~
     ${\mit\Delta} \alpha_{d}=0.022/\sqrt{\epsilon_{\ell}}$.
\hfill (\arabic{QQ})\addtocounter{QQ}{1}

This result is free from $m_H$ dependence since the Higgs-exchange
diagrams do not contribute to $\alpha_{\gamma1}$ and $\alpha_d$
determination within our approximation.
  \item[] \underline{$m_H=100$ GeV}
  \item[$\bullet$]
    $P_e=P_{\bar{e}}=-1, ~P_t = P_{\tilde{t}}=
     P_\gamma = P_{\tilde{\gamma}}=1/\sqrt{2}$,~
     $N_{\ell}\simeq 1.9\times 10^4 \epsilon_{\ell}$\\
     ${\mit\Delta} \alpha_{h1}=0.034/\sqrt{\epsilon_{\ell}}$,~
     ${\mit\Delta} \alpha_{d}=0.017/\sqrt{\epsilon_{\ell}}$.
\hfill (\arabic{QQ})\addtocounter{QQ}{1}
  \item[] \underline{$m_H=300$ GeV}
  \item[$\bullet$]
    $P_e=P_{\bar{e}}=-1,
     ~P_t = P_{\tilde{t}}=0,~
     P_\gamma = P_{\tilde{\gamma}}=1$,~
     $N_{\ell}\simeq 2.4\times 10^4 \epsilon_{\ell}$\\
     ${\mit\Delta} \alpha_{h1}=0.013/\sqrt{\epsilon_{\ell}}$,
     ${\mit\Delta} \alpha_{d}=0.015/\sqrt{\epsilon_{\ell}}$.
\hfill (\arabic{QQ})\addtocounter{QQ}{1}
  \item[] \underline{$m_H=500$ GeV}
  \item[$\bullet$]
    $P_e=P_{\bar{e}}=-1,~P_t = P_{\tilde{t}}=0,~
    P_\gamma = P_{\tilde{\gamma}}=1$,~
    $N_{\ell}\simeq 2.4\times 10^4 \epsilon_{\ell}$\\
     ${\mit\Delta} \alpha_{h1}=0.023/\sqrt{\epsilon_{\ell}}$,
     ${\mit\Delta} \alpha_{d}=0.015/\sqrt{\epsilon_{\ell}}$.~
\hfill (\arabic{QQ})\addtocounter{QQ}{1}
\item[$\bullet$]
    $P_e=P_{\bar{e}}=-1,~P_t = P_{\tilde{t}}=0,~
     P_\gamma = P_{\tilde{\gamma}}=1$,~ 
     $N_{\ell}\simeq 2.4\times 10^4 \epsilon_{\ell}$\\
     ${\mit\Delta} \alpha_{h2}=0.030/\sqrt{\epsilon_{\ell}}$,
     ${\mit\Delta} \alpha_{d}=0.015/\sqrt{\epsilon_{\ell}}$.~
\hfill (\arabic{QQ})\addtocounter{QQ}{1}
\end{description}

\vspace*{0.5cm}
\noindent
\underline{\underline{Final bottom-quark detection}}
\begin{description}
  \item[] \underline{$m_H=100$ GeV}
  \item[$\bullet$]
    $P_e=P_{\bar{e}}=-1,~P_t = P_{\tilde{t}}=
     -P_\gamma = P_{\tilde{\gamma}}=1/\sqrt{2}$,~ 
     $N_{b}\simeq 4.2\times 10^4 \epsilon_{b}$\\
     ${\mit\Delta} \alpha_{h1}=0.058/\sqrt{\epsilon_{b}}$,
     ${\mit\Delta} \alpha_{d}=0.026/\sqrt{\epsilon_{b}}$.
\hfill (\arabic{QQ})\addtocounter{QQ}{1}
  \item[] \underline{$m_H=300$ GeV}
  \item[$\bullet$]
    $P_e=P_{\bar{e}}=-1, ~P_t = P_{\tilde{t}}=
    -P_\gamma = P_{\tilde{\gamma}}=1/\sqrt{2},~ 
     N_{b}\simeq 4.2\times 10^4 \epsilon_{b}$\\
     \hspace*{-0.2cm}${\mit\Delta} \alpha_{h1}=0.009/\sqrt{\epsilon_{b}}$,
     ${\mit\Delta} \alpha_{h2}=0.074/\sqrt{\epsilon_{b}}$.
\hfill (\arabic{QQ})\addtocounter{QQ}{1}
  \item[$\bullet$]
    $P_e=P_{\bar{e}}=1,~P_t = P_{\tilde{t}}=
     -P_\gamma = P_{\tilde{\gamma}}=1/\sqrt{2}$,~
    $N_{b}\simeq 4.2\times 10^4 \epsilon_{b}$\\
     ${\mit\Delta} \alpha_{h1}=0.025/\sqrt{\epsilon_{b}}$,
     ${\mit\Delta} \alpha_{d}=0.019/\sqrt{\epsilon_{b}}$.
\hfill (\arabic{QQ})\addtocounter{QQ}{1}
 \item[$\bullet$]
    $P_e=P_{\bar{e}}=1,~P_t = P_{\tilde{t}}=
     P_\gamma = -P_{\tilde{\gamma}}=1/\sqrt{2}$,~
    $N_{b}\simeq 4.2\times 10^4 \epsilon_{b}$\\
     ${\mit\Delta} \alpha_{h2}=0.065/\sqrt{\epsilon_{b}}$,
     ${\mit\Delta} \alpha_{d}=0.010/\sqrt{\epsilon_{b}}$.~
\hfill (\arabic{QQ})\addtocounter{QQ}{1}
  \item[] \underline{$m_H=500$ GeV}
  \item[$\bullet$]
    $P_e=P_{\bar{e}}=-1,~P_t = P_{\tilde{t}}=1,~
     P_\gamma = P_{\tilde{\gamma}}=0$,~
    $N_{b}\simeq 4.6\times 10^4 \epsilon_{b}$\\
     ${\mit\Delta} \alpha_{h1}=0.030/\sqrt{\epsilon_{b}}$,
     ${\mit\Delta} \alpha_{d}=0.018/\sqrt{\epsilon_{b}}$.~
\hfill (\arabic{QQ})\addtocounter{QQ}{1}
 \item[$\bullet$]
    $P_e=P_{\bar{e}}=-1,~P_t = P_{\tilde{t}}=1,~
     P_\gamma = P_{\tilde{\gamma}}=0$,~ 
    $N_{b}\simeq 4.6\times 10^4 \epsilon_{b}$\\
     ${\mit\Delta} \alpha_{h2}=0.028/\sqrt{\epsilon_{b}}$,
     ${\mit\Delta} \alpha_{d}=0.014/\sqrt{\epsilon_{b}}$.~
\hfill (\arabic{QQ})\setcounter{QQE}{\arabic{QQ}}
\end{description}
%
\setcounter{equation}{\arabic{QQ}}
Using these results one can find (for known $m_H$) the most
suitable polarization for a determination of a given pair of
coefficients.

Note that it is difficult to determine $\alpha_{\gamma 1}$
and $\alpha_{\gamma 2}$ together in this analysis. 
Although we have found some
stable solutions that would allow for a determination of
$\alpha_{\gamma 1}$ in the lepton analysis, which we did not
find in \cite{Grzadkowski:2003tf}, the expected
precision is rather low. Nevertheless this is telling us that
the use of purely linear polarization for the laser is crucial
for measuring $\alpha_{\gamma 1}$. Unfortunately,
the statistical uncertainty of $\alpha_{\gamma 2}$ is still
large even in this analysis, so we did not list it in the
solutions which gave us good statistical precisions. Therefore,
we have to look for other suitable processes to determine this
parameter, for a review see \cite{Atwood:2000tu}.

On the other hand,
it was found that there are many combinations of polarization
parameters that make uncertainties of $\alpha_{h1,h2}$ and
$\alpha_{d}$ relatively small. For instance, analyzing the
$b$-quark final state with the polarization
    $P_e=P_{\bar{e}}=-1$, $P_t = P_{\tilde{t}}=1/\sqrt{2}$,
    $P_\gamma = -P_{\tilde{\gamma}}=-1/\sqrt{2}$
enables us to probe the properties of Higgs bosons whose mass
is around 300~GeV through the determination of $\alpha_{h1}$
and $\alpha_{h2}$.

As mentioned, the results were obtained for
${\mit\Lambda}=$ 1 TeV. If one assumes the new-physics scale to
be ${\mit\Lambda}=\lambda$ TeV, then all the above
results (${\mit\Delta}\alpha_i$) are replaced with
${\mit\Delta}\alpha_i/\lambda^2$, which means that the right-hand
sides of eqs.(\arabic{QQF})--(\arabic{QQE}) giving
${\mit\Delta}\alpha_i$ are all multiplied by $\lambda^2$.\\

\sec{SUMMARY}
Studying $\gamma\gamma \to t\bar{t}\to \ell X/b X$ in detail,
we derived optimal beam polarizations that minimize uncertainties
in the determination of $t\bar{t}\gamma$, $tbW$ and $\gamma\gamma H$
coupling parameters. To perform this analysis, we have
applied the optimal-observable method to the final
lepton/$b$-quark angular and energy distributions. 

We showed a number of two-parameter solutions,
most of which allow for the $\gaga H$- and $tbW$-couplings
determination. The expected precision of the measurement
of the Higgs coupling is of the order of $10^{-2}$ (for the scale of
new physics ${\mit\Lambda}=1$ TeV).
This shows that the $\gamma\gamma$ collider is going
to be useful for testing the Higgs sector of the SM.

Let us consider the top-quark-coupling determination in an
ideal case such that the beam polarizations could be easily
tuned and that the energy is sufficient for the on-shell
Higgs boson production. Then the best strategy would be to adjust
polarizations to construct semi-monochromatic $\gaga$ beams
such that $\sqrt{s_{\gaga}}\simeq m_H$ and on-shell Higgs
bosons are produced. This would allow for precise $\alpha_{h1,h2}$
measurement, so the virtual Higgs effects in $\gaga\to\ttbar$
would be calculable. Unfortunately,
as we have shown earlier, it is difficult to measure
$\alpha_{\gamma2}$ by looking just at $\ell X/b X$ final states
from $\gaga\to\ttbar$.
Therefore to fix $\alpha_{\gamma2}$, one should, e.g., measure the
asymmetries adopted in \cite{Choi:1995kp} to determine
the top-quark electric dipole moment which is proportional to
$\alpha_{\gamma2}$.
Then, following the analysis presented here,
one can determine $\alpha_{\gamma1}$ and $\alpha_d$. 

Finally, one must not forget that it is necessary to take into account
carefully the Standard Model contribution with radiative
corrections when trying to determine the anomalous couplings
in a fully realistic analysis. In particular this is significant
when we are interested in $C\!P$-conserving couplings since
the SM contributions there are not suppressed unlike the
$C\!P$-violating terms. On this subject, see for instance
\cite{Brandenburg:2005uu}.
 
\vspace{0.6cm}
\centerline{ACKNOWLEDGMENTS}
\vspace{0.3cm}
This work is supported in part by the State Committee for
Scientific Research (Poland) under grant 1~P03B~078~26 in
the period 2004--2006, the Grant-in-Aid for Scientific
Research No.13135219 and No.16540258 from the Japan
Society for the Promotion of Science, and the Grant-in-Aid
for Young Scientists No.17740157 from the Ministry of
Education, Culture, Sports, Science and Technology of Japan
and the  U. S. Department of Energy under Grant No.
DE-FG03-94ER40837.

\vspace*{1.3cm}

\end{document}